\documentclass[submission,copyright,creativecommons]{eptcs}
\providecommand{\event}{GandALF 2013} % Name of the event you are submitting to
\usepackage{breakurl}             % Not needed if you use pdflatex only.
\usepackage{amssymb}
\usepackage{amsmath}
\usepackage{pst-tree}
\usepackage{graphicx}
\usepackage{algorithm2e}

\usepackage{pstricks}
\usepackage[all]{xy}

\newcommand{\mnp}{
}

\newcommand{\natn}{\mathbb{N}}
\newcommand{\truth}{\mbox{{\bf true}}}
\newcommand{\falsity}{\mbox{{\bf false}}}

\newcommand{\unt}{\;{U}}
\newcommand{\tom}{{X}}
\newcommand{\fut}{F}
\newcommand{\alw}{G}
\newcommand{\pos}{{E}}
\newcommand{\nec}{{A}}
\newcommand{\clos}{\mbox{{\bf cl}}\;}
\newcommand{\huephi}{H_{\phi}}
\newcommand{\colourphi}{C_{\phi}}
\newcommand{\power}[1]{{\mathcal{P}}( #1 )}

\newcommand{\rX}{\; r_X}
\newcommand{\rA}{\; r_A}
\newcommand{\RCX}{\; R_X}

\newcommand{\vb}{\; | \;}

\newcommand{\lang}{L}

\newenvironment{definition}{Definition. }{}
\newenvironment{lemma}{Lemma. }{}

\title{A Faster Tableau for CTL*}
\author{Mark Reynolds
\institute{School of Computer Science and Software Engineering,
The University of Western Australia}
\email{mark.reynolds@uwa.edu.au}
}
\def\titlerunning{Tableau for CTL*}
\def\authorrunning{Mark Reynolds}
\begin{document}
\maketitle

\begin{abstract}
There have been several recent suggestions for tableau systems for
deciding satisfiability in the practically important branching time
temporal logic known as CTL*.
In this paper we present a streamlined and more traditional tableau
approach built upon the author's earlier theoretical work.

Soundness and completeness results are proved. A prototype implementation
demonstrates the significantly improved performance of the new approach on
a range of test formulas.
We also see that it compares favourably  to state of the art, game and automata based
decision procedures.
\end{abstract}

% no keywords

%%%%%%%%%%%%%%%%%%%%%%%%%%%%%%%MINE-->

\section{Introduction}
\label{sec:intro}
CTL* \cite{+emsi,+emhal86}
is an expressive branching-time
temporal logic 
extending the standard
linear 
PLTL \cite{Pnu77}.
The main uses of CTL*
are
for developing and checking the
correctness of complex
reactive systems  \cite{Eme90}
and as a basis for
languages (like ATL*)
for reasoning about multi-agent systems 
\cite{DBLP:conf/lfcs/GorankoS09}.

Validity of
formulas of CTL* is known to be decidable
with an automata-based decision procedure
of deterministic double exponential
time complexity \cite{+emsi,EJ88,VaS85}.
There is also an axiomatization \cite{Rey:ctlstar}.
Long term interest in 
developing a tableau approach as well
has been because they
are often more suitable for automated reasoning,
can quickly build
models of satisfiable formulas
and are more human-readable.
Tableau-style elements 
have indeed appeared earlier in some model-checking
tools for CTL*
but 
tableau-based satisfiability
decision procedures have only
just started to be developed
\cite{Rey:startab,FLL10}.

Our CTL* tableau
is of the 
tree, or top-down, form.
To decide the validity of $\phi$,
we build a tree
labelled with finite
sets of sets of formulas
using ideas called hues and colours originally from
\cite{Rey:ctlstar}
and further developed
in \cite{Rey:startabFM,Rey:startab}.
The formulas
in the labels come from a closure set containing
only subformulas of the formula being decided,
and their negations.
Those earlier works proposed
a tableau in the form of a roughly tree-shaped
Hintikka-structure,
that is,
it utilised labels on nodes which were
built from maximally consistent subsets of 
the closure set.
Each formula or its negation had to be in 
each hue.
In this paper we make the whole 
system much more efficient
by showing how we only need to 
consider subformulas
which are relevant to the decision.

In the older papers
we identified two sorts of looping:
good looping allowed up-links in our tableau tree
while
bad looping 
showed that a branch was just getting longer and
longer in an indefinite way.
In this paper we
tackle only the good looping aspect
and 
leave bad looping for a follow-on paper.

A publicly available prototype implementation of the
approach here is available
and 
comparisons 
with existing state of the art systems,
and its Hintikka-style predecessor,
show that
we are achieving orders of magnitude
speed-ups across
a range of examples.
As with any other pure tableau system, though,
this one is better at deciding satisfiable formulas
rather than unsatisfiable ones.

In section~\ref{sec:synsem} we give
a formal definition of CTL*
before section~\ref{sec:huecol} defines some basic building block concepts.
Subsequent sections
introduce the tableau shape,
 contain an example,
look at a loop checking rule and show soundness.
Section~\ref{sec:build} presents the tableau construction rules
and then we show
completeness.
Complexity, implementation and comparison
issues are discussed
briefly in section~\ref{sec:complex}
before a conclusion.
There is a longer version of this paper
available as \cite{Rey:fasttablong}.

\mnp

\section{Syntax and Sematics}
\label{sec:synsem}

Fix a countable set $\cal L$
of atomic propositions.
A (transition)  structure is a triple $M=(S,R,g)$
where:\\
\begin{tabular}{ll}
S & is the non-empty set of {\em states}\\
R & is a total binary relation $\subseteq S \times S$
 i.e. for every $s \in S$, there is some $t \in S$ such that
    $(s,t) \in R$.\\
$g$ & $: S \rightarrow \power{\cal L}$ is a labelling of the
states with sets of atoms.\\
\end{tabular}\\

Formulas are defined along $\omega$-long sequences
of states.
A {\em fullpath} in $(S,R)$ is an infinite sequence
$\langle s_0, s_1, s_2, ... \rangle$ of states such that
for each $i$,  $(s_i, s_{i+1}) \in R$.
For the fullpath
$\sigma = \langle s_0, s_1, s_2, ...\rangle$,
and any $i \geq 0$, we write
$\sigma_i$ for the state $s_i$ and
$\sigma_{\geq i}$ for the fullpath
$\langle s_i, s_{i+1}, s_{i+2}, ... \rangle$.

The formulas of CTL* are built from the
atomic propositions in $\cal L$ recursively
using classical connectives $\neg$ and $\wedge$ as
well as the temporal connectives
$\tom$, $\unt$ and $\nec$.
We use the standard abbreviations, $\truth$, 
$\falsity$, $\vee$,
$\rightarrow$,  $\leftrightarrow$, 
$\fut \alpha \equiv \truth \unt \alpha$,
$\alw \alpha \equiv \neg \fut \neg \alpha$, 
and $\pos \alpha \equiv \neg \nec
\neg \alpha$.

Truth of formulas is evaluated at fullpaths
in structures.
We write $M, \sigma \models \alpha$ iff the
formula $\alpha$ is true of the fullpath $\sigma$
in the structure $M=(S,R,g)$.
This is defined recursively by:\\
\begin{tabular}{lll}
$M, \sigma \models p$ & iff & $p \in g(\sigma_0)$, any $p \in {\cal L}$\\
$M, \sigma \models \neg \alpha$ & iff &
$M, \sigma \not \models \alpha$\\
$M, \sigma \models \alpha \wedge \beta$ & iff &
$M, \sigma \models \alpha$ and 
$M, \sigma \models \beta$\\
$M, \sigma \models \tom \alpha$ & iff &
$M, \sigma_{\geq 1} \models \alpha$\\
$M, \sigma \models \alpha \unt \beta$ & iff &   
there is $i \geq 0$ such that
$M, \sigma_{\geq i} \models \beta$ and
for each $j$, if $0 \leq j < i$ then
$M, \sigma_{\geq j}  \models \alpha$\\ 
$M, \sigma \models \nec \alpha$ & iff &
for all fullpaths $\sigma'$ such that
$\sigma_0 = \sigma'_0$ we have
$M, \sigma' \models \alpha$\\
\end{tabular}

We say that $\alpha$ is {\em valid} in CTL*, iff
for all transition structures $M$, for all fullpaths $\sigma$ in $M$,
we have
$M,\sigma \models \alpha$.
Say $\alpha$ is {\em satisfiable} in CTL* iff
for some transition structure $M$ and for some fullpath $\sigma$ in $M$,
we have
$M,\sigma \models \alpha$.
Clearly $\alpha$ is satisfiable
iff $\neg \alpha$ is not valid.

\section{Hues, Colours and Hintikka Structures}
\label{sec:huecol}

Fix the formula $\phi$ whose
satisfiability we are interested in.
We write
$\psi \leq \phi$
if $\psi$ is a subformula
of $\phi$.
The length of $\phi$ is $|\phi|$.
The {\em closure set} for $\phi$ is $\clos \phi =
\{ \psi, \neg \psi \vb
\psi \leq \phi \}$.

\begin{definition}[MPC]
Say that $a \subseteq \clos \phi$
is {\em maximally propositionally consistent (MPC)}
for $\phi$
iff
for all $\alpha, \beta \in \clos \phi$,
M1) if $\beta= \neg \alpha$ then
($\beta \in a$ iff $\alpha \not \in a$); and M2)
if $\alpha \wedge \beta \in \clos \phi$ then
($\alpha \wedge \beta \in a$ iff
both $\alpha \in a$ and $\beta \in a$).
\end{definition}

The concepts of hues and colours
were originally invented in 
\cite{Rey:ctlstar}
but we use 
particular formal 
definitions as presented
in \cite{Rey:startabFM,Rey:startab,Rey:fasttablong}.
A hue is supposed
to 
capture (approximately) a set
of formulas
which could all hold
together of one fullpath.
\begin{definition}[hue]
$a \subseteq \clos \phi$ is a {\em hue} for $\phi$,
or $\phi$-hue, iff
all these conditions hold:\\
\begin{tabular}{ll}
H1) &
$a$ is MPC;\\
H2) & if $\alpha \unt \beta \in a$
and $\beta \not \in a$ then $\alpha \in a$;\\
H3) & if $\alpha \unt \beta \in (\clos \phi) \setminus a$
then $\beta \not \in a$;\\
H4) &
if $\nec \alpha \in a$
then $\alpha \in a$.\\
\end{tabular}\\
Further, let $\huephi$ be the set of hues of $\phi$.
\end{definition}

For example, if $\neg (\nec  \alw (p \rightarrow \pos \tom p) 
\rightarrow (p \rightarrow \pos \alw  p))$,
the example known as $\neg \theta_{12}$ in \cite{Rey:startab},
then here is a hue known as $h{38}$:
\[
\begin{array}{l} \{ 
\neg (\nec  \alw (p \rightarrow \pos \tom p) 
\rightarrow (p \rightarrow \pos \alw  p)),  
(\nec  \alw (p \rightarrow \pos \tom p) 
\wedge \neg (p \rightarrow \pos \alw  p)), \\ 
\nec  \alw (p \rightarrow \pos \tom p),  
\alw (p \rightarrow \pos \tom p ),  
\truth,  
\neg \neg  (p \rightarrow \pos \tom p ),  \\
(p \rightarrow \pos \tom p ),  
p,  
\neg \neg  \pos \tom p,  
\pos \tom p,  
\neg \neg \tom p,  
\tom p,\\  
\neg (p \rightarrow \pos \alw  p),  
(p \wedge \neg \pos \alw  p ),  
\neg  \pos \alw  p,  
\nec \neg \alw p,  
\neg \alw p,  
\fut \neg p,  
\neg \neg p
\}\\
\end{array}
\]

The usual temporal successor relation
plays a role in determining allowed steps in the tableau.
The relation $r_X$ is put between hues $a$ and $b$
if a fullpath $\sigma$ satisfying $a$
could have a one-step suffix $\sigma_{\geq 1}$
satisfying $b$:
\begin{definition}[$r_X$]
For hues $a$ and $b$,
put $a \rX\ b$ iff
the following four conditions all hold:\\
\begin{tabular}{ll}
R1) &
if
$X \alpha \in a$ then $\alpha \in b$;\\
R2) &
if 
$\neg X \alpha \in a$ then $\neg \alpha \in b$;\\
R3) &
if
$\alpha \unt \beta \in a$ and $\neg \beta \in a$ then
$\alpha \unt \beta \in b$; and\\
R4) &
if
$\neg (\alpha \unt \beta) \in a$ and $\alpha \in a$ then
$\neg (\alpha \unt \beta) \in b$.\\
\end{tabular}
\end{definition}

We also introduced an
equivalence relation aiming
to tell whether two
hues could correspond to
fullpaths
starting at the same state.
We just need the hues to agree
on atoms and on
universal path quantified formulas:
\begin{definition}[$\rA$]
For hues $a$ and $b$,
put $a \rA\ b$ iff
the following two conditions both hold:
A1)  for all $p \in {\cal L}$, $p \in a$ iff $p \in b$; and
A2) $\nec \alpha \in a$ iff $\nec \alpha \in b$.
\end{definition}

Now we move up from the level of hues
to the level of colours.
Could a set of hues be exactly
the hues corresponding to
all the fullpaths
starting at a particular state?
We would need each pair of hues to 
satisfy $\rA$ but we would also
need hues to be in the set 
to witness all the 
existential path quantifications:

\begin{definition}[colour]
Non-empty $c \subseteq \huephi$ is a {\em colour}
of $\phi$,
or $\phi$-colour, iff the following
two conditions hold.
For all $a,b \in c$,
C1)  $ a \rA\ b$; and
C2)
if $a \in c$ and $\neg \nec \alpha \in a$
then
 there is $b \in c$ such that $\neg \alpha \in b$.
Let $\colourphi$ be the set of colours of $\phi$.
\end{definition}

The formulas $\neg \tom p, \pos \tom p$ are both in ${h37}$,
another hue from the example in \cite{Rey:startab},
so $\{ h37 \}$
is not a colour.
However,
$\tom p \in {h38}$
witnesses the existential
path quantification
so $\{ h{37}, h{38} \}$ is a colour.

We define a successor relation
$\RCX$ between colours.
It is defined in terms of 
the successor relation
$\rX$ between
the component hues
and it will be used to define
the successor relation between
tableau nodes,
themselves
corresponding to
states in transition structures,
in terms of the colours which they exhibit.
Note that colours,
and tableau nodes, will,
in general, have a non-singleton range
of successors
and this relation $R_X$
just tells us whether one node
can be one of the
successors of another node.

\begin{definition}[$\RCX$]
For all $c,d \in \colourphi$,
put $c \RCX\ d$ iff
for all $b \in d$ there is $a \in c$ such that $a \rX\ b$.
\end{definition}

It is worth noting that colours and hues 
are induced by
actual transition structures.
We will need these concepts in our completeness proof.

\begin{definition}[actual $\phi$-hue]
Suppose $(S,R,g)$ is a transition structure.
If $\sigma$ is a fullpath through $(S,R)$ then
we say that 
$h= \{ \alpha \in \clos \phi \vb 
(S,R,g), \sigma \models \alpha \}$
is the
{\em actual ($\phi$-) hue}
of $\sigma$ in $(S,R,g)$.
\end{definition}

It is straightforward to see that 
this is a $\phi$-hue.
It is also easy to show that
along any fullpath $\sigma$,
the relation
$r_X$ holds between the actual hue
of $\sigma$
and 
the actual hue 
of its successor 
fullpath $\sigma_{\geq 1}$.

\begin{definition}[actual $\phi$-colour]
If $s \in S$ then
the set of all actual hues 
of all fullpaths through $(S,R)$
starting at $s$
is called the 
{\em actual ($\phi$-) colour}
of $s$ in $(S,R,g)$.
\end{definition}

Again, it is straightforward
to show that this is indeed a
$\phi$-colour
and also that $R_X$
holds between the actual colour
of any state and
the actual colour
of any of its successors.

\begin{figure}
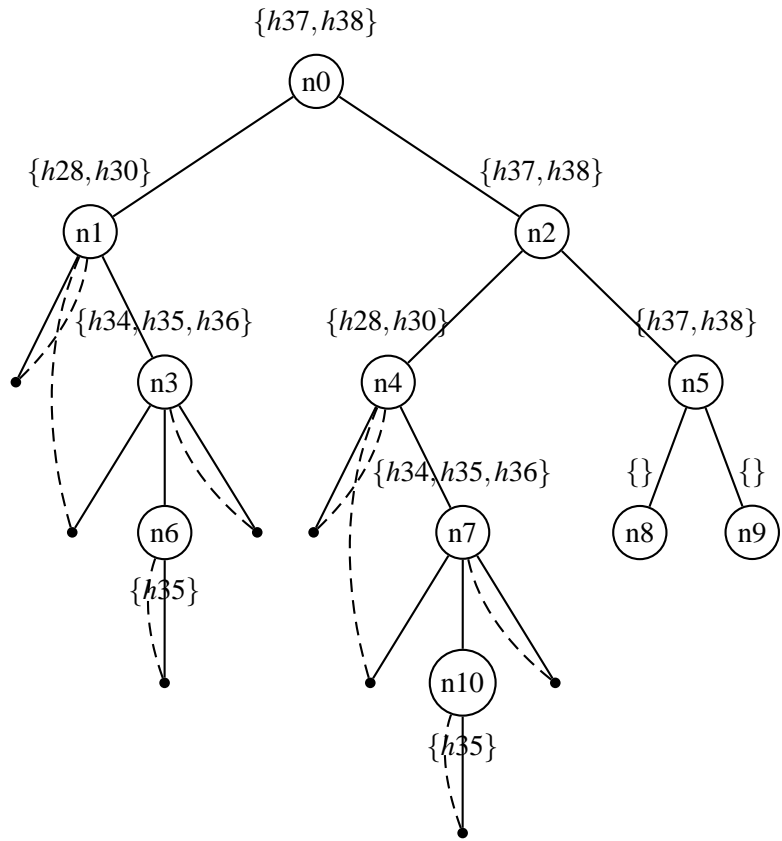

\begin{center}

\psset{radius=6pt,dotsize=4pt,treefit=loose}
\pstree[treemode=D]{\Tcircle[name=N0]{n0}~[tnpos=a]{$\{h37,h38\}$}}{%kids of 0							
    	\pstree{ \Tcircle[name=N1]{n1}~[tnpos=a]{$\{h28,h30\}$}}{%kids of 1
		\Tdot[name=N1c1]
		\pstree{\Tcircle[name=N3]{n3}~[tnpos=a]{$\{h34,h35,h36\}$}}{%kids of 3
			\Tdot[name=N3c1]
			\pstree{\Tcircle[name=N6]{n6}~[tnpos=b]{$\{h35\}$}}{%kids of 6
				\Tdot[name=N6c1]
}%end of kids of 6
			\Tdot[name=N3c3]
}%end of kids of 3
}%end kids of 1
	\pstree{\Tcircle[name=N2]{n2}~[tnpos=a]{$\{h37,h38\}$}}{%kids of 2
    		\pstree{ \Tcircle[name=N4]{n4}~[tnpos=a]{$\{h28,h30\}$}}{%kids of 4
			\Tdot[name=N4c1]
			\pstree{\Tcircle[name=N7]{n7}~[tnpos=a]{$\{h34,h35,h36\}$}}{%kids of 7
				\Tdot[name=N7c1]
				\pstree{\Tcircle[name=N10]{n10}~[tnpos=b]{$\{h35\}$}}{%kids of 10
					\Tdot[name=N10c1]
}%end of kids of 10
				\Tdot[name=N7c3]
}%end of kids of 7
}%end kids of 4
		\pstree{\Tcircle[name=N5]{n5}~[tnpos=a]{$\{h37,h38\}$}}{%kids of 5
    			\pstree{ \Tcircle[name=N8]{n8}~[tnpos=a]{$\{\}$}}{%kids of 8
}%end kids of 8
			\pstree{\Tcircle[name=N9]{n9}~[tnpos=a]{$\{\}$}}{%kids of 9
}%end kids of 9
}%end kids of 5
}%end kids of 2
}% end kids of 0
% Other connections
  \ncarc[linestyle=dashed,arcangle=-20]{N1c1}{N1}
  \ncarc[linestyle=dashed,arcangle=20]{N3c1}{N1}
  \ncarc[linestyle=dashed,arcangle=20]{N3c3}{N3}
  \ncarc[linestyle=dashed,arcangle=20]{N6c1}{N6}
  \ncarc[linestyle=dashed,arcangle=-20]{N4c1}{N4}
  \ncarc[linestyle=dashed,arcangle=20]{N7c1}{N4}
  \ncarc[linestyle=dashed,arcangle=20]{N7c3}{N7}
  \ncarc[linestyle=dashed,arcangle=20]{N10c1}{N10}
  
\end{center}
\caption{A Partial Tableau for $\neg \theta_{12}$}
\label{fig:egn12}
\end{figure}

\section{Tableau}
\label{sec:tableau}
\newcommand{\rot}{\mbox{\bf root}}

The tableaux we construct will be 
roughly tree-shaped:
the traditional upside down tree
with a root at the top,
predecessors and ancestors above,
successors and descendants below.
However, we will allow
up-links from a node
to one of its ancestors.
Each node will be labelled with a 
finite sequence of sets of formulas from the closure set.
We will call such a sequence of sets a
{\em proto-colour} or {\em pcolour}.
The sets, or {\em proto-hues (phues)}, in the pcolour
are ordered
and once completed
the node will have one (ordered) successor
for each phue.

The ordering of the successors
will match the ordering of the hues
(H3.1 and H6)
so that we know there
is a successor node
containing a successor phue
for each phue in the label.
The respective orderings
are otherwise arbitrary.

A {\em proto-hue (phue)} is just a subset of $\clos \phi$.

See Figure~\ref{fig:tab}
for our definition of a tableau.

\begin{figure}
\begin{definition}
A {\em tableau} for $\phi \in \lang$
is a tuple
$(T, s, \eta,\pi)$ such that:\\
\begin{tabular}{ll}
H1) &  $T$ is a non-empty set of {\em nodes}; one distinguished element called the {\em root};\\
H2) & $\eta$ is the phue label enumerator, so that  for each $t \in T$, $\eta_t: \natn \rightarrow 2^{\clos{\phi}}$ is a partial map, \\
H2.1) & the domain of $\eta_t$ is $\{ 0, 1, ..., n-1 \}$ for some $n>0$ denoted $|\eta_t|$;\\
 H2.2)  & $\eta_t(i)$ is the $i$th label phue of $t$ (if defined);\\
H3) & $s$ is the successor enumerator, so that
      for each $t \in T$, $s_t: \natn \rightarrow T$ is a partial map,\\
H3.1) &
the domain of $s_t$ is a subset of $\{ 0, 1, ..., |\eta_t|-1 \}$;
  $s_t(i)$ the $i$th successor of $t$;\\
 H3.3) & for each $t \in T$, there is a unique finite sequence $r_0, r_1, ..., r_k$ from $T$ 
 called the {\em ancestors} of $t$\\&
            such that  the $r_i$ are all distinct,
            $r_0$ is the root, $r_k=t$ and for each $j$, $r_{j+1}$ is a successor of $r_j$;\\
%H2.4) & $t$ is called a {\em leaf} iff it has no successors;\\ 
H4) & $\phi \in \eta_{\rot}(0)$;\\
H5) & $\pi$ is the predecessor map whereby if $t,u \in T$ then either $\pi^t_u$ is undefined\\&
and we say that $t$ is not a predecessor of $u$; or for all $j < |u|$, $\pi^t_u(j)=i<|t|$ and\\
& we say that the $i$th phue in $t$ is a predecessor of the
$j$ th hue in $u$.\\
H6) & if $s_t(i)=u$ then 
$\pi^t_u(0)=i$ (i.e. the $i$th phue in $t$ is a predecessor of the $0$th phue in $s_t(i)$);\\
\end{tabular}

\end{definition}
\caption{Definition of Tableau}
\label{fig:tab}

\end{figure}

\begin{definition}
Say that the tableau $(T,s,\eta,\pi)$
has {\em supported labelling}
if each formula in each phue in each label
is supported, as follows.
Consider a formula $\alpha \in \eta_t(i)$.
Determining whether $\alpha$ is support for not depends on the form
of $\alpha$:

\begin{tabular}{ll}
$-$ &
$p$ is supported in $\eta_t(0)$.
Otherwise, i.e. for $i>0$, it is only supported if $p \in \eta_t(0)$.\\
$-$ &
Same with $\neg p$.\\
$-$ &
$\neg \neg \alpha$ supported iff
$\alpha \in \eta_t(i)$.\\
$-$ &
$\alpha \wedge \beta$ supported iff
$\alpha \in \eta_t(i)$ and
$\beta \in \eta_t(i)$.\\
$-$ &
$\neg ( \alpha \wedge \beta )$ supported iff either
$\neg \alpha \in \eta_t(i)$ or
$\neg \beta \in \eta_t(i)$.\\
$-$ &
$X \alpha \in \eta_t(i)$ supported iff
1) there is $u \in T$ with $u=s_t(i)$
and
2)
for all $u \in T$, for all $j$ with\\& $\pi^t_u(j)=i$,
$\alpha \in \eta_u(j)$.\\
$-$ &
$\neg X \alpha \in \eta_t(i)$ supported iff
1) there is $u \in T$ with $u=s_t(i)$ and
2)
for all $u \in T$, for all $j$ with\\&
 $\pi^t_u(j)=i$,
$\neg \alpha \in \eta_u(j)$.\\
$-$ &
$\alpha U \beta \in \eta_t(i)$ supported iff
1) $\beta \in \eta_t(i)$; or
2) all 
2.1) $\alpha \in \eta_t(i)$;
2.2) there is $u \in T$ with\\& $u=s_t(i)$; and
 2.3)
for all $u \in T$, for all $j$ with $\pi^t_u(j)=i$,
$\alpha U \beta \in \eta_u(j)$.\\
$-$ &
$\neg (\alpha U \beta) \in \eta_t(i)$ supported iff
1) $\neg \beta \in \eta_t(i)$;
and 2) either
2.1) $\neg \alpha \in \eta_t(i)$; or
2.2) both 2.2.1)\\ & there is $u \in T$ with $u=s_t(i)$; and
2.2.2) for all $u \in T$, for all $j$ with $\pi^t_u(j)=i$,
$\neg( \alpha U \beta) \in \eta_u(j)$.\\
$-$ &
$A \alpha  \in \eta_t(i)$ supported iff
for all $j < |\eta_t|$,
$\alpha \in \eta_t(j)$.\\
$-$ &
$\neg A \alpha \in \eta_t(i)$ supported iff
there is some $j < |\eta_t|$,
$\neg \alpha \in \eta_t(j)$.\\
\end{tabular}

\end{definition}

A tableau is {\em successfully finished}
iff it 
has no leaves,
the predecessor relation is defined on all phues
and
the tableau does not fail any of the
three checks that we introduce below:
LG, NTP and
the non-existence of direct contradictions
(or $\falsity$)
in phues.

It is common, in proving properties
of tableau-theoretic approaches
to reasoning,
to refer to labelled
structures as {\em Hintikka structures}
if the labels 
are maximally complete
(relative to a closure set).
We say that one of our tableaux
$(T,s,\eta,\pi)$ is a Hintikka tableau
iff the
elements of each $\eta_t$
are all hues
(not just any phues).
The older tableau approach
in \cite{Rey:startab}
was based on Hintikka tableaux.

\section{Tableau Examples}
\label{sec:tegs}

Figure~\ref{fig:egn12}
is an example (unfinished) tableau 
illustrating general shape.
There are 11 nodes, 
each with successors marked,
and each labeled with a set of phues.
Note that some of the successor relations involve up-links:
$n1$ is a successor of $n3$.
We just name the phues rather than listing their contents.
There are more details about this example
in \cite{Rey:startab} as, in fact, it is
a Hintikka-tableau,
which is a special type of the tableau
we are introducing in this paper.
We use Hintikka-tableaux later in the
completeness proof here.

\begin{figure}
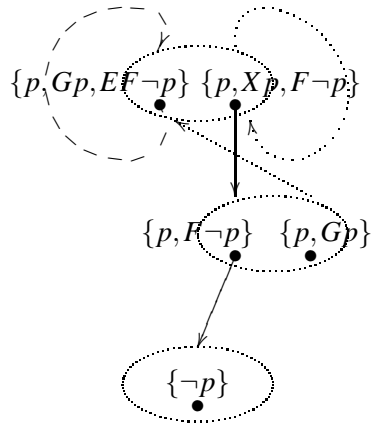


\[
\xy
(20,50)*\xycircle(10,5){.};
(7,50)*{\{p, Gp, EF\neg p\}};
(15,47)*{\bullet};
{\ar @{--} @/^/(15,47)*{};(12,40)*{}};
{\ar@{--} @/^/(9,40)*{};(0,48)*{}};
{\ar@{--} @/^/(0,52)*{};(9,60)*{}};
{\ar@{-->} @/^/(12,60)*{};(15,55)*{}};
{\ar@{.} @/^/(25,55)*{};(30,60)*{}};
{\ar@{.} @/^/(30,60)*{};(40,50)*{}};
{\ar@{.} @/^/(40,50)*{};(34,40)*{}};
{\ar@{..>} @/^/(34,40)*{};(27,45)*{}};
(31,50)*{\{p, Xp, F \neg p \}};
(25,47)*{\bullet};
{\ar(25,47)*{};(25,35)*{}};
(30,30)*\xycircle(10,5){.};
(20,30)*{\{p,F \neg p\}};
(25,27)*{\bullet};
{\ar@{..>} (35,35)*{};(17,45)*{}};
(37,30)*{\{p, G p \}};
(35,27)*{\bullet};
{\ar(25,27)*{};(20,15)*{}};
(20,10)*\xycircle(10,5){.};
(20,10)*{\{\neg p\}};
(20,7)*{\bullet};
\endxy
\]

\caption{Example tableau.}

\label{fig:meg}
\end{figure}

Figure~\ref{fig:meg}
shows a smaller tableau in more detail.
He we show the phues, which make up the pcolour labels
of nodes
and we show the predecessor or traceback map
in some cases.

\section{The LG test and Soundness}
\label{sec:lg}

In this section we will 
briefly describe the LG rule which is a tableau construction rule 
that prevents bad up-links being added.
LG
is used to test and possibly fail a tableau.
The test is designed to be used 
soon after any new up-link is
added after being proposed by the LOOP rule.
If the new tableau fails the LG test  then ``undo" the up-link and 
continue with alternative choices.
We then show that if a tableau finishes,
that is has no leaves, and passes the LG test
then it guarantees satisfiability.

There was also a very similar LG test in 
the earlier work on the original slower tableau method
\cite{Rey:startab}.
In that paper,
we show how to carry out the LG check 
on a tableau and we prove some results about
its use. 
The check is very much like a model check on the tableau so far.
We make sure that 
every phue in a label {\em matches},
or is a subset of an actual hue at that node
in a transition structure defined using
a valuation of atoms based on the labels.
It has polynomial running time in the size of the tableau so it is not a 
significant overhead on the overall tableau construction algorithm.

\begin{figure}
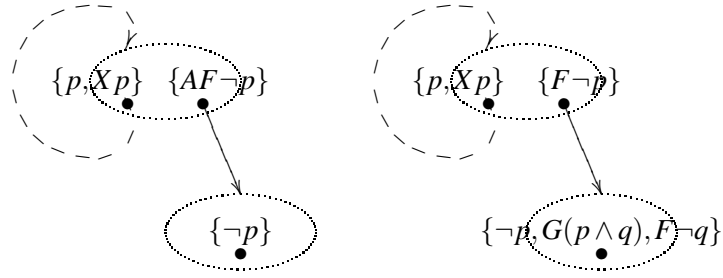


\[
\begin{array}{lll}
\xy
(20,50)*\xycircle(10,5){.};
(11,50)*{\{p, Xp\}};
(15,47)*{\bullet};
{\ar @{--} @/^/(15,47)*{};(12,40)*{}};
{\ar@{--} @/^/(9,40)*{};(0,48)*{}};
{\ar@{--} @/^/(0,52)*{};(9,60)*{}};
{\ar@{-->} @/^/(12,60)*{};(15,55)*{}};
(27,50)*{\{AF \neg p \}};
(25,47)*{\bullet};
{\ar(25,47)*{};(30,35)*{}};
(30,30)*\xycircle(10,5){.};
(30,30)*{\{ \neg p\}};
(30,27)*{\bullet};
\endxy
& \mbox{        } &
\xy
(20,50)*\xycircle(10,5){.};
(11,50)*{\{p, Xp\}};
(15,47)*{\bullet};
{\ar @{--} @/^/(15,47)*{};(12,40)*{}};
{\ar@{--} @/^/(9,40)*{};(0,48)*{}};
{\ar@{--} @/^/(0,52)*{};(9,60)*{}};
{\ar@{-->} @/^/(12,60)*{};(15,55)*{}};
(27,50)*{\{F \neg p \}};
(25,47)*{\bullet};
{\ar(25,47)*{};(30,35)*{}};
(30,30)*\xycircle(10,5){.};
(30,30)*{\{ \neg p, \alw (p \wedge q), \fut \neg q\}};
(30,27)*{\bullet};
\endxy
\end{array}
\]

\caption{LG examples: left fails LG; right
passes but eventually does not succeed}

\label{fig:lgnaeg}
\end{figure}

Due to space restrictions we do not go through the full
details of the only very slightly 
different LG rule used for the
faster tableaux here.
Instead we give some brief motivation examples.
The first example shows us that not all up-links are allowable:
e.g., a node labelled with $p, \nec \fut \neg p$ 
which also has an immediate loop.
See left hand example in Figure~\ref{fig:lgnaeg}.
The up-link would not be allowed by the LG rule.

The right hand example in Figure~\ref{fig:lgnaeg},
with an allowable up-link and also separately
an unsatisfiable leaf,
is allowed by LG.

\begin{figure}
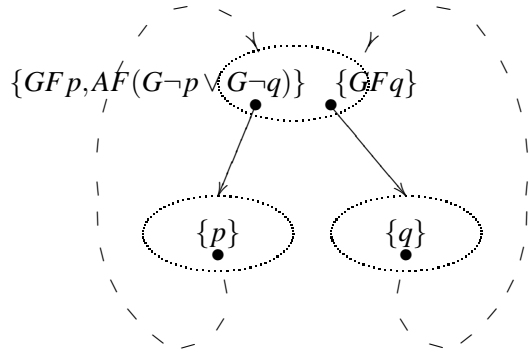


\[
\xy
(20,50)*\xycircle(10,5){.};
(2,50)*{\{GFp, \nec \fut ( \alw \neg p \vee \alw \neg q)\}};
(15,47)*{\bullet};
(31,50)*{\{GF q \}};
(25,47)*{\bullet};
{\ar(15,47)*{};(10,35)*{}};
(10,30)*\xycircle(10,5){.};
(10,30)*{\{ p\}};
(10,27)*{\bullet};
{\ar @{--} @/^/(11,23)*{};(6,15)*{}};
{\ar@{--} @/^/(2,16)*{};(-6,32)*{}};
{\ar@{--} @/^/(-6,36)*{};(0,57)*{}};
{\ar@{-->} @/^/(5,60)*{};(15,55)*{}};
{\ar(25,47)*{};(35,35)*{}};
(35,30)*\xycircle(10,5){.};
(35,30)*{\{ q \}};
(35,27)*{\bullet};
{\ar @{--} @/_/(34,23)*{};(39,15)*{}};
{\ar@{--} @/_/(43,16)*{};(51,32)*{}};
{\ar@{--} @/_/(51,36)*{};(45,57)*{}};
{\ar@{-->} @/_/(40,60)*{};(30,55)*{}};
\endxy
\]

\caption{These two loops fail LG.}

\label{fig:lgtwoeg}
\end{figure}

The example in Figure~\ref{fig:lgtwoeg} has two loops, each 
one individually acceptable but not both.
The LG rule fails the tableau when both up-links are added.

Now we show that if $\phi$ has a successfully finished tableau
then $\phi$ is satisfiable.
This is the soundness Lemma.

\begin{lemma}
If $\phi$ has a successfully finished tableau
then $\phi$ is satisfiable.
\end{lemma}

Here we just outline the proof: details in \cite{Rey:fasttablong}.
Say that $(T,s,\eta,\pi)$ is a successfully finished tableau for $\phi$.
Define a structure $M=(T,R,g)$ 
by interpreting the $s$ relation as a transition relation $g$,
and using $\eta$ to define the valuation $g$ on nodes.

 By definition of matching,
 after a final check of LG
 there is some actual hue $b$ of the root 
 such that $\eta_{\rot}(0) \subseteq b$.
 This means that $\phi$ holds
 along some fullpath in the 
 final structure.

\section{Building a tree}
\label{sec:build}

In this section we briefly describe how a tableau is built via some simple operations, or rules.
We start with an initial tree of one root node labelled with just one
phue containing only $\phi$.
The rules allow formulas to be added inside hues in labels,
new hues to be added in labels
and
new nodes to be added as successors 
of existing nodes.
The rules are generally non-deterministic allowing
a finite range of options,
or choices,
at any application.

There are some properties to check such as LG,
described above, and
NTP described below.
We also check that there are no hues containing 
both a formula and its negation,
and we check that $\falsity$ is not contained in a phue.
If these checks fail then the tableau has
failed and we will need to backtrack 
to explore other possible options 
at choice points along the way.

The tableau succeeds if there are no leaves.

\subsection{Basic Tableau Rules}
\label{sec:rules}

Here are most of the basic rules, in an abbreviated notation:

\noindent
\begin{picture}(400,100)
\put(0,80){2NEG: $\frac{\{\{\neg \neg \alpha\}\}}{\{\{\alpha\}\}}$}
\put(80,80){CONJ: $\frac{\{\{\alpha \wedge \beta\}\}}{\{\{\alpha, \beta\}\}}$}
\put(160,80){DIS: $\frac{\{\{\neg (\alpha \wedge \beta) \}\}}{\{\{\neg \alpha\} \} \vb \{\{\neg \beta\}\}}$}
\put(260,80){NEX: $\frac{\{\{X \alpha\}\} \rightarrow \{\{\}\}}{\{\{X \alpha\}\} \rightarrow \{\{\alpha\}\}}$}
\put(350,80){NNX: $\frac{\{\{\neg X \alpha\}\} \rightarrow \{\{\}\}}{\{\{\neg X \alpha\}\} \rightarrow \{\{\neg \alpha\}\}}$}
\put(0,50){UNT: $\frac{\{\{ \alpha U \beta \}\} \rightarrow \{\{\}\}}{
\{\{\alpha U \beta, \beta \}\} \rightarrow \{\} \vb
\{\{\alpha U \beta, \alpha \}\} \rightarrow \{\{\alpha U \beta \}\}}$}
\put(200,50){NUN: $\frac{\{\{ \neg (\alpha U \beta) \}\} \rightarrow \{\{\}\}}{
\{\{\neg (\alpha U \beta), \neg \beta, \neg \alpha \}\} \rightarrow \{\} \vb
\{\{\neg (\alpha U \beta), \neg \beta, \alpha \}\} \rightarrow \{\{ \neg (\alpha U \beta) \}\}}$}
\put(0,20){ATM: $\frac{\{\{ p \}, \{ \} \}}{\{\{p \},  \{p \}\} }$}
\put(80,20){NAT: $\frac{\{\{ \neg p \}, \{ \} \}}{\{\{ \neg p \},  \{ \neg p \}\} }$}
\put(190,20){POS: $\frac{\{\{ \neg A \alpha \} \}}{
\{\{\neg A \alpha, \neg \alpha \}\} \vb \{\{\neg A \alpha \}, \{ \neg \alpha \} \}}$}
\put(330,20){NEC: $\frac{\{\{ A \alpha \}, \{ \} \}}{
\{\{A \alpha, \alpha \},  \{ \alpha \}\} }$}
\end{picture}

The rules are described in detail in \cite{Rey:fasttablong}
but the notation gives the main ideas.
Here are details of a few of the rules above.

{\bf DIS:}
If $\neg (\alpha \wedge \beta) \in \eta_t(j)$ 
then 
can extend $(T,s,\eta,\pi)$ to $(T',s',\eta',\pi')$
via either:
DIS1 or DIS2 as follows.
DIS1 produces
$(T',s',\eta',\pi')$
such that
$T'=T$,
$s'=s$,
and
for all $t' \neq t$,
$\eta_{t'}=\eta_t$
and
for all $i' \neq i$,
$\eta'_t(i')=\eta_t(i')$.
However,
$\eta'_t(i)=\eta_t(i) \cup \{ \neg \alpha \}$.
DIS2 is similar but use $\beta$ instead of $\alpha$.

{\bf NEX:}
If $X \alpha  \in \eta_t(i)$ 
and
there is $u \in T$ and $j$
with $\pi^t_u(j) =i$
then can extend $(T,s,\eta,\pi)$ to $(T',s',\eta',\pi')$
such that
$T'=T$,
$s'=s$,
and
$\eta'_{u}(j)=\eta_u(j) \cup \{ \alpha \}$.
If $t \in T$ 
but there is no $s_t(j) \in T$
then extend $(T,s,\eta,\pi)$ to $(T',s',\eta',\pi')$
using new object $t^+$
such that
$T'=T \cup \{ t^+ \}$,
$s'_t(i)=t^+$,
$\eta'_{t^+}(0)=\{  \}$
and $\pi'^t_{t^+}(0)=i$.
For all other arguments, $s'$, $\eta'$ and $\pi'$ inherit values from
$s, \eta$ and $\pi$ respectively.

{\bf ATM:}
If an atom $p  \in \eta_t(j)$ 
and $k<|\eta_t|$
then can extend $(T,s,\eta,\pi)$ to $(T',s',\eta',\pi')$
such that
$T'=T$,
$s'=s$,
and
for all $t' \neq t$,
$\eta_{t'}=\eta_t$
and
for all $i' \neq k$,
$\eta'_t(i')=\eta_t(i')$.
However,
$\eta'_{t}(k)=\eta_t(k) \cup \{ p \}$.

{\bf POS:}
If $\neg A \alpha  \in \eta_t(j)$
and $n=|\eta_t|$ 
then can extend $(T,s,\eta,\pi)$ to $(T',s',\eta',\pi')$
via one of
$\mbox{POS}_k$ for some
$k=0, 1, 2, ..., n$
as follows.
For $k<n$, $\mbox{POS}_k$ involves 
extending $(T,s,\eta,\pi)$ to $(T',s',\eta',\pi')$
where
$T'=T$,
$s'=s$,
and
for all $t' \neq t$,
$\eta_{t'}=\eta_t$
and
for all $i' \neq k$,
$\eta'_t(i')=\eta_t(i')$.
However,
$\eta'_{t}(k)=\eta_t(k) \cup \{ \neg \alpha \}$.
However, $\mbox{POS}_n$ involves 
extending $(T,s,\eta,\pi)$ to $(T',s',\eta',\pi')$
where
$T'=T$,
$s'=s$,
and
for all $t' \neq t$,
$\eta_{t'}=\eta_t$
and
for all $i' \neq k$,
$\eta'_t(i')=\eta_t(i')$.
However,
$\eta'_{t}(k)=\eta_t(k) \cup \{ \neg \alpha \}$.

There are also a couple of rules not sketched above.

{\bf PRED:}
If $t,u \in T$ and 
$u$ is a successor of $t$ but
$\pi(t_u(j))$ is not defined
then we can extend $(T,s,\eta,\pi)$ to $(T',s',\eta',\pi')$
via one of
$\mbox{PRED}_k$ for some
$k=0, 1, 2, ..., |\eta_t|-1$
as follows.

For $k<|\eta_t|$, $\mbox{PRED}_k$ involves 
extending $(T,s,\eta,\pi)$ to $(T',s',\eta',\pi')$
where
$T'=T$,
$s'=s$,
and
$\eta'=\eta$.
However,
$\pi'^t_{u}(j)=k$.

For $k=|\eta_t|$,
$\mbox{PRED}_k$ involves 
extending $(T,s,\eta,\pi)$ to $(T',s',\eta',\pi')$
where
$T'=T$, but
$\eta'=\eta$ but giving $t$ an extra empty phue $\eta'_{t}(k)=\{\}$;
and
$s=s'$.

Later we need to add a $k$th successor for $t$ and
fill in formulas in $\eta'_t(k)$.

Note that $t$ now potentially becomes 
unsupported, untraceable and unfinished, again.

{\bf LOOP:}
Suppose $t$ is an ancestor of the parent $u^-$ of $u$,
 then we can choose either to replace the $u^-$ to $u$ edge by an up-link from $u^-$ to $t$, or to not do that replacement (and continue the branch normally).

(It is worth remembering which choice you make and not try that again if it did not work.)

Note that, as in normal successors, we will also put
$s_{u^-}(i)=t$ and $\pi^{u^-}_t(0)=i$
where 
previously we had
$s_{u^-}(i)=u$.
All the other phues in $\eta_t$ will also have to have
predecessors chosen amongst the phues in
$\eta_{u^-}$. We will use the PRED rule to do this for each one.

Note also that making such an up-link can possibly cause
a subsequent consequential failure of the tableau.
A contradiction could be introduced into the hues of $t$,
the NTP could fail and/or the LG property could fail.
It is possible to test for a few of these potential problems
just before making use of this rule and act accordingly.

\subsection{The NTP check: nominated thread property}

The LG property check that
every looping path 
is noticed by the
labels in nodes.
The converse requirement is taken care
of by the much simpler NTP check.

We put a
special significance on the initial
hue in each colour label.
This, along with the next condition,
helps us ensure that each hue
actually has a fullpath
witnessing it.
We are going to require the following 
property, NTP,
of the tableaux which we construct.

First some auxiliary definitions:
\begin{definition}[hue thread]
Suppose $\sigma$ is a path through $(T,s,\eta,\pi)$.
A {\em hue thread} through $\sigma$
is
a sequence $\xi$ of hues such that
$|\xi|=|\sigma|$,
for each $j<|\xi|$, $\xi_j \in \eta(\sigma_j)$
and for each $j<|\xi|-1$,
$\xi_j r_X \xi_{j+1}$.
\end{definition}

\begin{definition}[fulfilling hue thread]
Suppose $\sigma$ is a path through  $(T,s,\eta,\pi)$
and $\xi$ is a  hue thread through $\sigma$.
We say that $\xi$ is fulfilling iff
either $|\sigma|<\omega$,
or 
 $|\sigma|=\omega$ and all the eventualities in each
$\xi_i$ are witnessed
by some later $\xi_j$;
i.e.
if $\alpha \unt \beta \in \xi_i$
then there is $j \geq i$
such that
$\beta \in \xi_j$.
\end{definition}

\begin{definition}[the nominated thread property]
We say that the tableau $(T,s,\eta,\pi)$
has the {\em nominated thread property}
(NTP) iff the following holds.
Suppose that
for all $t \in T$ such that $0<|s_t|$, $s_t(0)$ is an
ancestor of $t$
and that $t_0=s_t(0), t_1, ..., t_k=t$
is a non-repeating sequence with each
 $t_{j+1}=s_{t_j}(0)$.
 Let $\sigma$ be the fullpath
 $\langle t_0, t_1, ..., t_k, t_0, t_1, ..., t_k, t_0, t_1, ... \rangle$
 and
 $\xi$ be the sequence
 $
 \langle \eta_{t_0}(0), \eta_{t_1}(0), ..., \eta_{t_k}(0), 
\eta_{t_0}(0), ... \rangle
 $
 of hues in $\sigma$.
Then $\xi$ is a fulfilling hue thread for $\sigma$.
\end{definition}

It is straightforward to prove that this
is equivalent to checking
that
each eventuality in $\eta_{t_0}(0)$
(or in all, or any, $\eta_{t_i}(0)$)
is witnessed in at least one of the $\eta_{t_j}(0)$.
So it is neither hard 
to implement nor computationally complex.

Using the rules described above,
using any applicable one at any stage,
allows construction
of 
tableaux.
We know that the LG rule ensures that 
any successful ones which we build thus
will guarantee that $\phi$ is
satisfiable.
In the next section we consider
whether we can build a successful tableau
for any satisfiable formula
in the way.

\section{Completeness Using the Hintikka Tableau}
\label{sec:completeness}

In \cite{Rey:startab}, the completeness result for the tableau
in that paper, shows that for
any satisfiable CTL* formula
 there is a finite model 
satisfying certain useful properties 
and 
from that we can find a successful tableau (as defined in that paper)
for the formula.
In fact the tableau constructed in that paper is just a special
form of the tableaux that we are constructing in this paper:
they are Hintikka structures.

\begin{definition}
\label{def:sht}
A structure $(T,s,\eta,\pi)$ is a
{\em Standard Hintikka Tableau} for $\phi$
iff
$(T,s,\eta,\pi)$ is a finite
finished successful tableau for $\phi$
and
for each $t$,
for each $i$,
$\eta_t(i)$ is an MPC subset 
of $\clos(\phi)$.
\end{definition}

Thus, in a Hintikka tableau, the
labels tell us exactly which formulas
hold there.

The completeness result in \cite{Rey:startab}
shows the following, in terms of the concepts defined
in this paper:

\begin{lemma}
If $\phi \in L$ is satisfiable
then it has a
Standard Hintikka Tableau.
\end{lemma}

The proof of this lemma is a straightforward
translation of the definitions from \cite{Rey:startab}
but we need to specify how to define our current predecessor relation
$\pi$ and we also need to
check that the tableau is finished.

The predecessor relation $\pi$ is not made explicit in the tableau
structures of the earlier paper.
Instead we require that the
colour of  a node $t$ is related by a
successor relation $R_X$ between colours
to the colour of any successor $t'$.
This means that for any hue in the colour of $t'$
there is a hue $h$ in the colour of $t$
such that $h$ and $h'$ are related by
a successor relation between hues.
We can use such a hue $h$ as the predecessor of $h'$
and so define $\pi$.

To show that the tableau $(T,s,\eta,\pi)$
is finished,
we just need to check 
all the rules of our tableau construction
and make sure none
require the tableau to be changed in any way.
This needs to be done each rule at a time,
and needs to be done carefully,
although it is straightforward.

The proof in \cite{Rey:startab}
uses a finite model theorem
for CTL* 
to obtain a {\em branch boundedness}
result on the Hintikka tableau.
We can guarantee existence of a such a tableau
with a certain function of the length
of the formula bounding  the length of each branch
(before an up-link).
The bound is triple exponential in the
length of the formula, so rather large.

Thus we can conclude that each satisfiable formula
has a tableau,
but we can not yet claim that it
is a tableau which can be constructed
by our rules.

In the rest of this section we 
describe how
we can show that if $\phi$ is satisfiable
then there is a sequence 
of applications of our tableau rules 
that allow the construction of 
a successful tableau for $\phi$.
Suppose $\phi$ is satisfiable.
From the lemma above we
know that there is a successful, branch-bounded, supported
tableau $T^{- \infty}=(T',s',\eta',\pi')$ for 
$\phi$.

In \cite{Rey:fasttablong}, we show how to build a related, successful tableau
for $\phi$
in a step by step manner
only using the 
construction rules from section~\ref{sec:rules}.
Thus we make a sequence
$T^0, T^1, ...$ of tableaux 
each one using a construction step to get to the next.

In order to use $T^{- \infty}$ to guide us,
we also construct
a sequence of maps
$w_0, w_1, w_2, ...$,
each $w_i$ relating the 
phues of the labels of the nodes of $T^i$ to the hues of the 
labels of the nodes
of $T^{- \infty}$.

Thus each $w_i$ maps ordered pairs which are
nodes paired with indices
to other such pairs.
Suppose that $T^i=(T,s,\eta,\pi)$
and $T^{-\infty}=(T',s',\eta',\pi')$.
Say $t \in T^i$ and 
$j < |\eta_t|$.
Then $w_i(t,j)$ will be defined: say that
$w_i(t,j)=(u,k)$ for $u \in T'$.
Then $k<|\eta'_u|$.
The idea in this example is that $w_i$ is associating the $j$th phue
of $t$ with the $k$th phue of $u$.

All the while during the
construction
we ensure that 
 $w_i$ maps 
each node in $T^i$ to a node in $T^{-\infty}$
which has a superset label.

We also show that the constructed tableau 
does not fail at any stage if one of the check rules
such as LG, NTP or the
existence of direct contradictions in phues.
This follows from the fact that the
phues in its labels are subsets 
of the hues in the
labels of the Hintikka tableau.

If $T$ is finished (leafless), supported and 
all predecessors exist then we are done.
If $T$ is not supported then choose any
formula $\alpha$ in any phue in the label
of any node that is not supported.
Depending on the form of $\alpha$
we apply one of the tableau rules
to add some successor,
or some phue
and/or some formula(s)
in a phue that will ensure that
$\alpha$ is then supported.
See \cite{Rey:fasttablong} for details.

There are only a finite number of 
formulas that can be added in hues in labels in a finite structure which is a subset of 
$T^{-\infty}$.
This guarantees
that the process will eventually terminate.

Thus every satisfiable formula
has a successful tableau
which can be found via our set
of rules.

In fact, we can go further and get an
even better completeness result.
We can show that each formula
$\phi$ only has a
finite number of tableaux
which respect the branch bounds
and a simple bound on branching factor.
Furthermore, if there is a successful
tableau then there will be one
obeying these bounds.
There are at most $2^{|\phi|}$ hues
and so each node
in a Hintikka tableau has at most
$2^{|\phi|}$ successors:
by the form of completeness proof we can
enforce the same bound on our
more general tableaux.
As we also have a finite bound on the length
of branches
there are clearly only finitely many
tableaux for any particular
$\phi$.

\begin{lemma}
Given $\phi$,
there are only a finite 
number of tableaux 
which respect the branch length bound 
and the branching degree bounds.
\end{lemma}

In this definition of tableau we have guaranteed
termination of any 
tableau construction
algorithm
by putting a simple 
but excessive bound
on the length of branches.
This allows us to 
conclude failure in a finite time
and to also abbreviate the
search for successful tableaux.

\section{Stopping Repetition: coming up in follow-on paper}
\label{sec:stop}

In this paper we have only briefly
mentioned the limit on the length
of branches as a way of guaranteeing that
there are only finitely many tableau,
and so that a search will terminate
one way or another.
The limit,
based on a theoretical upper bound
on the minimal CTL* model size,
is very generous and hence this is an inefficient
way of
cutting short tableau searches.
Being so generous slows down both negative and positive
satisfiability reports.

In order to make some sort of working
implementation to demonstrate the 
practicality of this tableau
it is necessary to have a better way
of preventing the construction
of wastefully long branches.
For want of better terminology
we will call such a facility,
a ``repetition checker".

The task of making a quick
and more generally usable
repetition checker will be left 
to be advanced and presented at a later date.
In fact, eventually we hope to
provide a useful  set of criteria
for earlier termination
of construction of branches
depending on the 
properties of the
sequence of colours
so far.
A simple example
of the sort of criterion 
is the repeated appearance of the same
sequence of colours and hues
along a non-branching path
without being able to construct any
up-links.
Other more sophisticated ideas
are easily suggested
but we want to
develop a more systematic
set of tests 
before presenting this
in future work.

In \cite{Rey:startab}, we
present some basic repetition checking 
tests
for the Hintikka style tableau.
These can be used in order to allow
some faster automated tableau
construction.
The tests can be modified to work with 
our sparser labels,
and we will present full details in a future paper.
There are many opportunities for more thorough 
repetition checks as well.

\section{Complexity, Implementation and Comparisons}
\label{sec:complex}

Say that $| \phi |=l$.
Thus $\phi$ has $\leq l$ subformulas
and $\clos \phi$ contains
at most $2l$ formulas.
Since each hue contains,
for each $\alpha \leq \phi$ at most
one of $\alpha$ or $\neg \alpha$,
there are at most
$\leq 2^l$ hues.
Thus there are
less than $2^{2^l}$ colours.
It is straightforward to see that there
is a triple exponential upper bound
if the tableau search algorithm
uses the double exponential bound on branch length
\cite{Rey:startab}
to curtail searches down long branches.

A prototype implementation written by the author
shows that for many interesting, albeit relatively small,
formulas, the experimental performance
of the system is relatively impressive.
There are some preliminary results 
detailed in
\cite{Rey:fasttablong}
which show a comparison of running times
with the older Hintikka-style tableau technique
of \cite{Rey:startab}
and the 
state of the art game-based CTL* reasoner from
 \cite{FLL10}.
 In general the new reasoner is more than an order
 of magnitude quicker at deciding
 formulas
 from a range of basic and distinctive
 CTL* validities and their negations
 and a few other satisfiable formulas.
 The implementation is available as Java code
 for public download \cite{Rey:fasttablong}.
 Online reasoner coming soon.

The implementation
for the new technique that is used in these experiments,
uses some basic repetition checking
derived from the
checks given earlier in the Hintikka-style system \cite{Rey:startab}.
The new, slightly modified
versions of these mechanisms are not described in the current paper.
Instead they will be described in a future paper.

In \cite{FLL10},
four series of formulas are suggested to 
examine asymptotic behaviour.
Timing results for our system 
on these formulas are
presented in Table~\ref{fig:fllcomp}.
We compare
the performance of our new tableau with
the state of the art in game-based
techniques for deciding CTL*.
This is using published performance
of the reasoner from \cite{FLL10}
as reported in experiments 
in \cite{DBLP:journals/entcs/McCabe-Dansted11}.
Consider the following series of formulas:
$\alpha_1=AFGq$,
$\beta_1=AFAGq$
and for each $i \geq 1$,
$\alpha_{i+1}=AFG\alpha_i$
and
$\beta_{i+1}=AFAG \beta_i$.
In table~\ref{fig:fllcomp},
we compare the performance of the 
Hintikka-style tableau system from
\cite{Rey:startab},
the game-based reasoner
from \cite{FLL10} and
the new tableau system of this paper
(using basic repetition checking)
on the growing series built from these formulas.
Although the running times, are on different computers,
and so not directly comparable,
we can see the difference in asymptotic performance.
Running times greater than an hour or two are curtailed.
From the results we see that
we have achieved 
very noticeable and significant improvements
in performance
on the satisfiable examples.

Pure tableau-style reasoning on unsatisfiable formulas
often involves
exhaustive searches and
the new technique is not immune to
such problems.
See the 400 series of examples in the asymptotic
experiments.
We will say more about these examples
when proposing some new 
repetition mechanisms
in the future.

\begin{figure}
{\small
\begin{tabular}{lllllll}
\hline
\# & formula  & length &  sat? & MRH & FLL & NEW\\
 &   &  &   & \cite{Rey:startab} & \cite{FLL10} & this paper\\
\hline
\hline
101 & $\alpha_1 \rightarrow \beta_1$ & 20  & Y & 330 & 120 & 39 \\
\hline
102 & $\alpha_2 \rightarrow \beta_2$ & 35  & Y & $>10^5$ & 130 &  43\\
\hline
103 & $\alpha_3 \rightarrow \beta_3$ & 50  & Y & out of time & 120 & 69 \\
\hline
108 & $\alpha_8 \rightarrow \beta_8$ & 125  & Y & out of time & 380 & 664 \\
\hline
113 & $\alpha_{13} \rightarrow \beta_{13}$ & 200  & Y & out of time & $>10^5$ & 2677 \\
\hline
115 & $\alpha_{15} \rightarrow \beta_{15}$ & 230  & Y & out of time & $>10^6$ & 4228 \\
\hline
119 & $\alpha_{19} \rightarrow \beta_{19}$ & 290  & Y & out of time & out of time  & 9468\\                             
\hline
\hline
201 & $\neg (\alpha_1 \rightarrow \beta_1)$ & 21  & Y & 350 & 120 & 172 \\
\hline
202 & $\neg (\alpha_2 \rightarrow \beta_2)$ & 36  & Y & $>10^5$ & 170 &  117\\
\hline
203 & $\neg (\alpha_3 \rightarrow \beta_3)$ & 51  & Y & out of time & 2270 & 213 \\
\hline
204 & $\neg ( \alpha_4 \rightarrow \beta_4)$ & 66  & Y & out of time & $> 10^6$ & 377\\
\hline
205 & $\neg ( \alpha_5 \rightarrow \beta_5)$ & 81  & Y & out of time & out of time & 673 \\
\hline
212 & $\neg ( \alpha_{12} \rightarrow \beta_{12})$ & 186  & Y & out of time &out of time & 7153 \\                   
\hline
\hline
301 & $\beta_1 \rightarrow \alpha_1$ & 20  & Y & 340 & 130 & 48 \\
\hline
302 & $\beta_2 \rightarrow \alpha_2$ & 35  & Y & $>10^5$ & 140 &  50\\
\hline
303 & $\beta_3 \rightarrow \alpha_3$ & 50  & Y & out of time & 140 & 86 \\
\hline
312& $ \beta_{12} \rightarrow \alpha_{12}$ & 185  & Y & out of time & 30970 & 3333\\
\hline
314 & $ \beta_{14} \rightarrow \alpha_{14}$ & 215  & Y & out of time & $>10^6$ & 5512\\           
\hline
316 & $ \beta_{16} \rightarrow \alpha_{16}$ & 245  & Y & out of time & out of time & 8627 \\   
\hline
319 & $ \beta_{19} \rightarrow \alpha_{19}$ & 290  & Y & out of time & out of time &  15615 \\           
\hline
\hline
401 & $\neg (\beta_1 \rightarrow \alpha_1)$ & 21  & N & 400 & 760 & 1801 \\
\hline
402 & $\neg (\beta_2 \rightarrow \alpha_2)$ & 36  & N & $>10^5$ & 48670 & $>10^5$ \\
\hline
403 & $\neg (\beta_3 \rightarrow \alpha_3)$ & 51  & N & out of time & $>10^6$ & out of time \\
\hline
\end{tabular}
}%end small
\caption{Asymptotic Examples: Running Times (milliseconds)}
\label{fig:fllcomp}
\end{figure}

There are some, more theoretical descriptions of 
other game-based and automata-based
techniques for model-checking CTL*
in older papers such as
\cite{Lange00modelchecking},
\cite{DBLP:conf/cav/BernholtzVW94}
and
\cite{DBLP:conf/focs/KupfermanV05}.
However, these do not seem 
directly applicable to satisfiability decisions
and/or there do not seem to be any easily publicly available
implemented tools based on these
approaches.

\section{Conclusion}
\label{sec:concl}

In this paper we have presented, albeit in a fairly high level sketch, a traditional tableau
approach to reasoning with the important logic CTL*.
Soundness and completeness results are proved
and prototype implementation
demonstrates the significantly improved performance of the new approach on
a range of test formulas.

The next task in this direction is
to build on the foundation here and
present full details and proofs of the
repetition checking mechanisms that
can be used with the tableau construction.
There are some basic repetition mechanisms available in the previous, Hintikka style 
tableau
\cite{Rey:startab} but they need to be modified slightly. 
There are opportunities for additional techniques.
It is also important to
improve and document the
rule-choice algorithms which
have a bearing on
running times.

In the future, 
it will be useful to develop reasoning tools which combine
the latest 
in tableaux, automata and game-based
approaches to CTL*.
Having tools working in parallel should allow faster decisions.
It will also be useful to extend the work to 
logics of multi-agent systems such as ATL* 
and strategy logic \cite{DBLP:conf/concur/MogaveroMPV12}.

%\nocite{*}
\bibliographystyle{eptcs}

\end{document}